\documentstyle[12pt]{article}
\newlength{\extraspace}
\newlength{\extraspaces}
\newcommand{\be}{\begin{equation}
\addtolength{\abovedisplayskip}{\extraspaces}
\addtolength{\belowdisplayskip}{\extraspaces}
\addtolength{\abovedisplayshortskip}{\extraspace}
\addtolength{\belowdisplayshortskip}{\extraspace}}
\newcommand{\ee}{\end{equation}}
\newcommand{\ba}{\begin{eqnarray}
\addtolength{\abovedisplayskip}{\extraspaces}
\addtolength{\belowdisplayskip}{\extraspaces}
\addtolength{\abovedisplayshortskip}{\extraspace}
\addtolength{\belowdisplayshortskip}{\extraspace}}
\newcommand{\ea}{\end{eqnarray}}
\newcommand{\newsection}[1]{
\vspace{7mm}
\pagebreak[3]
\addtocounter{section}{1}
\setcounter{equation}{0}
\setcounter{subsection}{0}
\setcounter{footnote}{0}
\begin{center}
{\large {\bf \thesection. #1}}
\end{center}
\nopagebreak
\medskip
\nopagebreak
\hspace{3mm}}
\newcommand{\nonu}{\nonumber \\[.5mm]}

\newcommand{\e}{\, {\rm e}}

\newcommand{\VEV}[1]{\left\langle {#1} \right\rangle}

\newcommand{\figureone}{
\begin{figure}[tb]
\setlength{\unitlength}{1mm}
\begin{picture}(100, 55)(-25, 20)
\put(50, 50){\thicklines\circle{15}}
\put(57, 50){\thicklines\line(1, 0){15}}
\put(47, 56.5){\thicklines\line(-3, 5){7}}
\put(46, 44){\thicklines\line(-3, -5){7}}
\put(55.5, 54.5){\line(6, 5){12}}
\put(55.5, 45.5){\line(6, -5){12}}
\put(51.5, 57){\line(1, 5){3}}
\put(51.5, 43){\line(1, -5){3}}
\put(43.5, 52.5){\line(-5, 2){14}}
\put(43.5, 47.5){\line(-5, -2){14}}
\put(30.5, 50){\circle*{0.5}}
\put(31, 53){\circle*{0.5}}
\put(31, 47){\circle*{0.5}}
\put(57, 68){\circle*{0.5}}
\put(57, 32){\circle*{0.5}}
\put(60, 67){\circle*{0.5}}
\put(60, 33){\circle*{0.5}}
\put(62.5, 65){\circle*{0.5}}
\put(62.5, 35){\circle*{0.5}}
\put(37, 71){${\bf 1}$}
\put(37, 27){${\bf 2}$}
\put(75, 49){${\bf 3}$}
\put(22, 49){$m$}
\put(62, 71){$n$}
\put(62, 26){$l$}
\end{picture}
\caption{The configuration of the operators for $I_{lmn}$.}
\end{figure}}
\newcommand{\figuretwo}{
\begin{figure}[tb]
\setlength{\unitlength}{1mm}
\begin{picture}(100, 60)(0, 15)
\put(40, 50){\thicklines\circle{15}}
\put(35, 55){\thicklines\line(-5, 4){11}}
\put(34.3, 45.8){\thicklines\line(-5, -4){11}}
\put(33.2, 52){\thicklines\line(-3, 1){12.5}}
\put(45.5, 54.5){\thicklines\line(5, 4){11}}
\put(46, 46){\thicklines\line(5, -4){11}}
\put(47, 51.5){\thicklines\line(3, 1){12.5}}
\put(23, 50.5){\circle*{0.5}}
\put(23.5, 46.5){\circle*{0.5}}
\put(25, 43){\circle*{0.5}}
\put(57, 50.5){\circle*{0.5}}
\put(56.5, 46.5){\circle*{0.5}}
\put(55, 43){\circle*{0.5}}
\put(18.5, 64.5){$+$}
\put(15, 56){$+$}
\put(18.5, 33.5){$+$}
\put(58.5, 65){$-$}
\put(62, 56){$-$}
\put(59, 33.5){$-$}
\put(37.5, 16){(a)}
\put(110, 50){\thicklines\circle{15}}
\put(117, 50){\thicklines\line(1, 0){15}}
\put(102.7, 50){\thicklines\line(-1, 0){15}}
\put(115, 55){\thicklines\line(1, 1){10.5}}
\put(115, 45){\thicklines\line(1, -1){10.5}}
\put(110, 57){\thicklines\line(0, 1){14}}
\put(110, 43){\thicklines\line(0, -1){14}}
\put(105, 55){\thicklines\line(-1, 1){10.5}}
\put(105, 45){\thicklines\line(-1, -1){10.5}}
\put(135, 49){$-$}
\put(128, 66.5){$+$}
\put(127.5, 30.5){$-$}
\put(108.5, 74.5){$+$}
\put(108.5, 24){$+$}
\put(89, 66.5){$+$}
\put(89.5, 31){$+$}
\put(81.8, 49){$-$}
\put(107.5, 16){(b)}
\end{picture}
\caption{Tachyon amplitudes. (a) Non-vanishing amplitudes.
(b) An example of vanishing amplitudes.}
\end{figure}}
\newcommand{\figurethree}{
\begin{figure}[tb]
\setlength{\unitlength}{1mm}
\begin{picture}(100, 55)(0, 20)
\put(55, 50){\thicklines\circle{15}}
\put(95, 50){\thicklines\circle{15}}
\put(62, 50){\thicklines\line(1, 0){26}}
\put(48, 50){\thicklines\line(-1, 0){15}}
\put(49.5, 54.5){\line(-5, 4){12}}
\put(49.5, 45.5){\line(-5, -4){12}}
\put(54, 57){\line(-1, 6){2.5}}
\put(54, 42.8){\line(-1, -6){2.5}}
\put(90.4, 55.4){\line(-5, 6){10}}
\put(90.4, 44.5){\line(-5, -6){10}}
\put(95, 57){\line(0, 1){15}}
\put(95, 42.8){\line(0, -1){15}}
\put(99, 56){\thicklines\line(3, 5){8}}
\put(99, 44){\thicklines\line(3, -5){8}}
\put(101.7, 52.5){\line(3, 1){14}}
\put(101.7, 47.5){\line(3, -1){14}}
\put(43, 64){\circle*{0.5}}
\put(43, 36){\circle*{0.5}}
\put(45.5, 66){\circle*{0.5}}
\put(45.5, 34){\circle*{0.5}}
\put(48.5, 67){\circle*{0.5}}
\put(48.5, 33){\circle*{0.5}}
\put(85.5, 66){\circle*{0.5}}
\put(85.5, 34){\circle*{0.5}}
\put(88.5, 67.5){\circle*{0.5}}
\put(88.5, 32.5){\circle*{0.5}}
\put(91.5, 68){\circle*{0.5}}
\put(91.5, 32){\circle*{0.5}}
\put(113, 53){\circle*{0.5}}
\put(113, 47){\circle*{0.5}}
\put(113.5, 50){\circle*{0.5}}
\put(27.5, 51.5){${\bf 1}$}
\put(27, 46){$+$}
\put(105.5, 24.5){${\bf 2}$}
\put(110, 27){$+$}
\put(105.5, 72.5){${\bf 3}$}
\put(110, 71){$-$}
\put(41, 70){$q$}
\put(41, 27){$k$}
\put(81.5, 73.5){$n \! - \! q$}
\put(81, 24.5){$m \! - \! k$}
\put(119.5, 48){$l$}
\put(70, 53){\vector(1, 0){10}}
\put(71.5, 56){$p^\mu$}
\put(76, 55.5){$-$}
\put(119, 57.5){$+$}
\put(119, 41){$+$}
\put(93.5, 22){$+$}
\put(75, 29){$+$}
\put(75, 69.5){$+$}
\put(93.5, 76){$+$}
\put(49.5, 22){$+$}
\put(49.5, 76){$+$}
\put(32, 33){$+$}
\put(32, 65){$+$}
\end{picture}
\caption{A factorization of the amplitude in the case (i).}
\end{figure}}
\newcommand{\figurefour}{
\begin{figure}[tb]
\setlength{\unitlength}{1mm}
\begin{picture}(100, 55)(0, 20)
\put(55, 50){\thicklines\circle{15}}
\put(95, 50){\thicklines\circle{15}}
\put(49.2, 45.8){\thicklines\line(-3, -2){12.4}}
\put(97.4, 43.2){\thicklines\line(2, -5){5.6}}
\put(101, 54){\thicklines\line(3, 2){12.6}}
\put(51.2, 55.8){\line(-2, 3){8.4}}
\put(48.1, 50.8){\line(-6, 1){14.7}}
\put(100.8, 45.8){\line(3, -2){12.6}}
\put(102, 50){\line(1, 0){15}}
\put(97.7, 56.5){\line(2, 5){5.5}}
\put(93.5, 56.8){\line(-1, 6){2.5}}
\put(62, 50){\thicklines\line(1, 0){26}}
\put(70, 53){\vector(1, 0){10}}
\put(71.5, 56){$p^\mu$}
\put(76, 55.5){$+$}
\put(41.8, 62.8){\circle*{0.5}}
\put(39.4, 59.8){\circle*{0.5}}
\put(37.7, 56.4){\circle*{0.5}}
\put(111.5, 41.8){\circle*{0.5}}
\put(112.7, 44.3){\circle*{0.5}}
\put(113.1, 47.1){\circle*{0.5}}
\put(99.4, 68){\circle*{0.5}}
\put(97, 68.5){\circle*{0.5}}
\put(94.5, 68.5){\circle*{0.5}}
\put(31.5, 36){${\bf 1}$}
\put(35, 32.5){$-$}
\put(101, 22.5){${\bf 2}$}
\put(105.5, 25){$-$}
\put(113.5, 66){${\bf 3}$}
\put(117, 62.5){$+$}
\put(33, 62){$q$}
\put(93.5, 74.5){$n \! - \! q$}
\put(118.5, 41.5){$l$}
\put(38, 70.5){$+$}
\put(27, 53){$+$}
\put(116.5, 34.5){$+$}
\put(121, 49){$+$}
\put(104.5, 73){$+$}
\put(87.5, 75){$+$}
\end{picture}
\caption{A factorization of the amplitude in the case (ii).}
\end{figure}}
\begin{document}
\addtolength{\baselineskip}{.7mm}
\thispagestyle{empty}
\begin{flushright}
STUPP--92--128 \\ March 1992
\end{flushright}
\vspace{.6cm}
\begin{center}
{\large{\bf{Disk Amplitudes in Two-Dimensional \\[2mm]
Open String Theories}}}
\footnote{A talk given at the Workshop on New Aspect of Quantum
Field Theory, Institute for Nuclear
Study, University of Tokyo, February 17-19, 1992} \\[20mm]
{\sc Yoshiaki Tanii} \\[7mm]
and \\[7mm]
{\sc Shun-ichi Yamaguchi} \\[12mm]
{\it Physics Department, Saitama University \\[2mm]
Urawa, Saitama 338, Japan} \\[20mm]
{\bf Abstract}\\[1cm]
{\parbox{13cm}{\hspace{5mm}
Disk amplitudes of tachyons in two-dimensional open string
theories (two-dimensional quantum gravity coupled to $c \leq 1$
conformal field theories) are obtained using the continuum
Liouville field approach. The structure of momentum singularities
is different from that of sphere amplitudes and is more
complicated. It can be understood by factorizations of the
amplitudes with the tachyon and the discrete states as
intermediate states.}}
\end{center}
\vfill
\newpage
\setcounter{section}{0}
\setcounter{equation}{0}
%
%
\newsection{Introduction}
Two-dimensional conformal field theories with central charge
$c \leq 1$ coupled to gravity have been extensively studied
recently using the continuum Liouville field
approach \cite{DDK}, \cite{SEPO}. They are useful as toy models to
understand four-dimensional quantum gravity and non-critical
string theories. When the $c \leq 1$ conformal field theories are
realized by a scalar field, they can be regarded as critical
string theories in two-dimensional target space with non-trivial
background fields. Amplitudes (correlation functions) of these
models have been computed in
refs.\ \cite{GTW}-\cite{KITA} and are shown to exhibit
characteristic momentum singularities. In the case of $c = 1$
theory on a sphere, the origin of the singularities has been
understood as short distance singularities arising from the
operator product expansion (OPE) of vertex
operators \cite{POLYAKOVSELF}, \cite{SATAFACT}, \cite{DFKU}.
The physical states of the model are tachyon and discrete
states \cite{POLYAKOVSELF}, \cite{KAC}. Both of them appear
as intermediate states of the factorized amplitude, which
arises by the OPE, and give rise to the pole singularities.
\par
In ref.\ \cite{DISK} we have studied two-dimensional conformal
field theories with central charge $c \leq 1$ coupled to gravity
on a disk. Surfaces with boundaries, such as the disk,
naturally appear in string theories including open strings in
addition to closed strings. In particular, we have computed disk
amplitudes of three open string tachyons. The structure of momentum
singularities is different from that of the sphere amplitudes and
is more complicated.
\par
In this paper we first review the results of ref.\ \cite{DISK}.
We then study the OPE and factorizations of the disk amplitudes.
We find that all momentum singularities are explained by the
factorizations with the tachyon and the discrete states as
intermediate states. This factorization analysis applies to
$c < 1$ theories using the $c < 1$ discrete states \cite{ITOH} as
well as to the $c = 1$ theory.
\par
Two-dimensional open string theories have been discussed also
by Bershadsky and Kutasov \cite{BEKU}. They gave results of the
general $N$-point disk amplitudes of open string tachyons.
They also discussed factorizations and momentum singularities of
the amplitudes. Our results in ref.\ \cite{DISK} and in this paper
are consistent with theirs.
\par
\newpage
%
%
\newsection{Tachyons and Discrete States}
We consider a two-dimensional conformal matter coupled to gravity
on a disk $D$. The matter conformal field theory is realized by a
scalar field $X$ with the background charge $\alpha_0 < 0$ and has
the central charge $c = 1 - 12 \alpha_0^2$. After the conformal
gauge fixing
$g_{\alpha\beta} = \e^{\alpha_+ \phi}\hat g_{\alpha\beta}$
with a fixed reference metric $\hat g_{\alpha\beta}$, the system
is described by the matter field $X$ and the Liouville field
$\phi$ with the action \cite{DDK}, \cite{DISK}
\ba
S[\hat g_{\alpha\beta}, X^\mu] &\!\!\! = &\!\!\!
{1 \over 8\pi} \int\nolimits_D d^2 z \sqrt{\hat g}
\left(\hat g^{\alpha\beta} \partial_\alpha X_\mu
\partial_\beta X^\mu - i \hat R Q_\mu X^\mu
+ 8 \mu \e^{\alpha_+ \phi} \right) \nonu
&\!\!\! &\!\!\! + \, {1 \over 4\pi} \int\nolimits_{\partial D}
d x \hat E \left(-i \hat k Q_\mu X^\mu
+ 4\lambda \e^{{1 \over 2} \alpha_+ \phi} \right),
\label{stringaction}
\ea
where $\hat R$ is the scalar curvature, $\hat k$ is the geodesic
curvature of the boundary $\partial D$ and $\hat E$ is the
one-dimensional metric on the boundary induced from
$\hat g_{\alpha\beta}$. The parameters $\mu$ and
$\lambda$ are renormalized values of the bulk and the boundary
cosmological constants respectively. We have used two-vector
notations $X^\mu = (X,\phi)$ and $Q^\mu=(-2 \alpha_0,-iQ)$, and
their inner products are defined by the Euclidean flat metric
$\delta_{\mu\nu}$. The disk can be conformally mapped onto the
upper-half complex plane
$\{ z \in {\bf C} \, | \, {\rm Im} \, z \ge 0 \}$. We use
$x = {\rm Re} \, z$ to parametrize the boundary
${\rm Im} \, z = 0$.
\par
The parameters $Q$ and $\alpha_+$ in eq.\ (\ref{stringaction}) are
fixed by requiring that the theory does not depend on the gauge
choice $\hat g_{\alpha\beta}$. Using the Neumann boundary condition
on both of the Liouville and the matter fields, we
obtain \cite{DDK}, \cite{DISK}
\be
Q = \sqrt{25 - c \over 3}, \quad
\alpha_\pm = - {1 \over 2\sqrt{3}}
\left( \sqrt{25-c} \mp \sqrt{1-c} \,\right),
\label{qalpha}
\ee
where we have introduced $\alpha_-$ for later use.
The action (\ref{stringaction}) with (\ref{qalpha}) describes a
conformal field theory with the central charge $26$. Therefore, we
can regard it as a critical open-closed string theory in
two-dimensional target space $X^\mu$ with non-trivial
background fields.
\par
We now discuss the physical operators of the theory, which are
primary fields of unit conformal weight.
Here we will consider only the open string vertex operators, which
represent emission and absorption of open strings at the boundary.
(For the closed string vertex operators, see
ref.\ \cite{DISK}.) The simplest such operator is the tachyon
operator
\be
O_{\rm o}^\pm(p)
= \int d x \hat E \,{\rm e}^{{1 \over 2}ip \cdot X },
\quad \beta_\pm(p) = - {Q \over 2} \pm (p - \alpha_0),
\label{boundaryop}
\ee
where $p^\mu = (p,-i \beta)$ and $p \cdot X = p_\mu X^\mu$. The
signs in $\beta_{\pm}$ are called the chirality of the tachyon.
Note that the boundary cosmological term operator in
eq.\ (\ref{stringaction}) is a particular case of the tachyon
vertex operator $O_{\rm o}^+ (p=0)$.
\par
At higher oscillator levels there exist nontrivial physical
operators only at discrete values of momenta. They are primary
fields for the discrete states. The discrete states in the $c=1$
theory \cite{KAC}, \cite{POLYAKOVSELF} are well-known and have
been extensively discussed recently.
The discrete states exist also in $c<1$ theories and can be
obtained from those in the $c=1$ theory by the SO(2, {\bf C})
transformation \cite{ITOH}.
The energy-momentum tensor of the $c<1$ theory can be
obtained from that of the $c=1$ theory by a complex
rotation in the two-dimensional space $X^\mu$
\be
X^\mu(c) = {\Omega^\mu}_\nu(c) \, X^\nu(c=1), \quad
{\Omega^\mu}_\nu(c) = {1 \over 2\sqrt{2}} \left( \matrix{
Q &\!\!\! -2i\alpha_0 \cr
2i\alpha_0 &\!\!\! Q \cr}\right).
\label{rotation}
\ee
The background charge $Q^\mu$ and two-momentum $p^\mu$ also rotate
by the matrix ${\Omega^\mu}_\nu (c)$.
In the $c=1$ theory the discrete states have momenta \cite{KAC}
\be
p_{r,\, t} = {r-t \over \sqrt{2}}, \quad
\beta^{(\pm)}_{r,\, t} = {-2 \pm (r+t) \over \sqrt{2}}
\label{discreteone}
\ee
at level $n=rt$, where $r, \; t$ are non-negative integers.
Momenta of the discrete states in the $c<1$ theory are obtained
from eq.\ (\ref{discreteone}) by the rotation (\ref{rotation})
\be
\left\{ \matrix{
p^{(+)}_{r,\, t} = {1 \over 2}
\left[ (t-1)\alpha_+ - (r-1)\alpha_- \right] \hfill \cr
\beta^{(+)}_{r,\, t} = - {1 \over 2}
\left[ (t-1)\alpha_+ + (r-1)\alpha_- \right] \hfill \cr} \right.,
\quad
\left\{ \matrix{
p^{(-)}_{r,\, t} = {1 \over 2}
\left[ -(r+1)\alpha_+ + (t+1)\alpha_- \right] \hfill \cr
\beta^{(-)}_{r,\, t} = {1 \over 2}
\left[ (r+1)\alpha_+ + (t+1)\alpha_- \right] \hfill \cr} \right..
\label{discrete}
\ee
The `energy' $\beta^{(\pm)}_{r,\, t}$ satisfies
$\beta^{(+)}_{r,\, t} > - {Q \over 2}$ and
$\beta^{(-)}_{r,\, t} < - {Q \over 2}$.
We call the discrete states with momenta $p^{(+)}_{r, t}$ and
$p^{(-)}_{r, t}$ as S-type and A-type respectively.
In contrast to the $c=1$ case (\ref{discreteone}), the
discrete states of S-type and the discrete states of A-type
have different values of momenta.
\par
\newpage
%
%
\newsection{Three-Point Tachyon Amplitudes}
In this section we compute three-point amplitudes
of the open string tachyons (\ref{boundaryop}) on the disk.
We consider the case $\mu = 0$ and $\lambda \not= 0$
in the action (\ref{stringaction}) for simplicity. One may consider
the case $\mu \not= 0$ and $\lambda = 0$ similarly.
When both of $\mu$ and $\lambda$ are non-zero, one can use a
perturbation expansion in one of them while treating the other
exactly. As in the case of the sphere amplitudes \cite{GTW} we can
first integrate over the zero modes
$X_0, \; \phi_0$ $(X = X_0 + \tilde X, \phi=\phi_0 + \tilde\phi)$
and obtain
\be
\VEV{ O_{\rm o}(p_1)\,O_{\rm o}(p_2)\,O_{\rm o}(p_3)}
= 2\pi \delta \left( \sum_{i=1}^3 p_i - 2 \alpha_0 \right)
    {4 \Gamma (-s) \over |\alpha_+|}
    \left( {\lambda \over \pi} \right)^s
    \tilde A (p_1,p_2,p_3),
\label{atilde}
\ee
\vspace{-7mm}
\ba
\tilde A (p_1, p_2, p_3) &\!\!\! = &\!\!\!
\; \int \prod_{i=1}^3 \left[ d x_i \hat E \right]
{1 \over V_{\rm SL(2, \, {\bf R})}} \nonu
&\!\!\! &\!\!\! \times \VEV{ \prod_{i=1}^3
\e^{{1 \over 2}i p_i \tilde X(x_i)} }_{\tilde X}
\VEV{ \prod_{i=1}^3 \e^{{1 \over 2} \beta_i \tilde\phi(x_i)}
\left( \int d x \hat E
\e^{{1 \over 2}\alpha_+ \tilde\phi} \right)^s }_{\tilde\phi},
\label{zeromodeint}
\ea
where $V_{\rm SL(2, \, {\bf R})}$ is the volume of the gauge group
SL(2, {\bf R}) generated by the conformal Killing vectors on the
disk. We have used shorthand notations $\beta_i = \beta(p_i)$ and
defined
\be
s = {1 \over |\alpha_+|} \left( Q + \sum_{i=1}^3 \beta_i \right).
\label{es}
\ee
The expectation values in eq.\ (\ref{zeromodeint}) are with
respect to the non-zero modes $\tilde X, \tilde\phi$, which have
a free action. When $s$ is a non-negative integer, we can
evaluate the expectation values. In this case one has to
interpret a singular factor in eq.\ (\ref{atilde}) as
$\Gamma (-s) \left( {\lambda \over \pi} \right)^s
\rightarrow (-1)^{s+1} {1 \over s!}
\left({\lambda \over \pi} \right)^s \ln \lambda$ \cite{DFKU}.
We will consider only the case with a non-negative integer $s$.
Fixing the SL(2, {\bf R}) gauge symmetry
by choosing the positions of the operators as
$x_1\!=\!0, x_2\!=\!1, x_3\!=\!\infty$, the $x$-integrations
can be evaluated by a technique similar to that of
ref.\ \cite{DF}. We obtain \cite{DISK}
\be
\tilde A(p_1, p_2, p_3)
= s! \sum_{l,m,n=0 \atop l+m+n=s}^s I_{lmn}(a, b, \rho),
\label{integralrep}
\ee
where $a \!=\! - \alpha_+ \beta_1, \ b\! =\! - \alpha_+ \beta_2,
\ \rho\! =\! -{1 \over 2} \alpha_+^2$.
The function $I_{lmn}$ denotes a contribution from an integration
region where $l$ `cosmological operators'
$\e^{\alpha_+ \tilde\phi/2}$ are inserted between operators 2
and 3, $m$ are inserted between operators 1 and 2, and $n$ are
inserted between operators 3 and 1 as shown in Fig.\ 1.
The explicit form of $I_{lmn}(a, b, \rho)$ can be found in
ref.\ \cite{DISK}.
\figureone
\par
We have two kinematical constraints on the momenta, i.e.
the momentum conservation in eq.\ (\ref{atilde}) and the definition
of $s$\ (\ref{es}) for a given $s$.
There are two possible choices of the chiralities
satisfying these constraints: (i) $(+, +, -)$ and (ii) $(-, -, +)$.
Without loosing generality we can choose the tachyon 3 to have
the opposite chirality to others.
By the kinematical constraints, only one momentum, e.g. $p_1$,
is independent and other momenta are given by
\ba
{\rm (i)} & &
p_2 = -p_1 - {1 \over 2}(s+2)\alpha_+ + {1 \over 2}\alpha_-, \quad
p_3 = {1 \over 2} s \alpha_+ + {1 \over 2}\alpha_-, \nonu
{\rm (ii)} & &
p_2 = -p_1 + {1 \over 2}(s-1)\alpha_+ + \alpha_-, \quad
p_3 = -{1 \over 2} (s+1) \alpha_+.
\label{region}
\ea
In both cases the momentum $p_3$ has a fixed value.
\par
Let us consider each case separately. In the case (i),
$I_{lmn}$ in eq.\ (\ref{integralrep}) is simplified to
\ba
I_{lmn} \,&\!\!\!
= &\!\!\! \,(-1)^{l+n} \,{\pi \over s!}
\left[ {\pi \over \Gamma (1+\rho)} \right]^s
\prod_{k=1}^l {1 \over \sin (\pi k \rho)} \,
\prod_{k=1}^n {1 \over \sin (\pi k \rho)} \nonu
&\!\!\! &\!\!\! \times \, {1 \over \Gamma (1 - \rho + a )
\, \Gamma ( (1-s)\rho - a )} \,
\prod_{k=0}^m {1 \over \sin [\pi ((1-n-k)\rho - a )]}.
\label{caseone}
\ea
It has poles at
\be
p_1 = - {1 \over 2}(n+k+1)\alpha_+ + {1 \over 2}N\alpha_-
\quad \ \ \ (\, k=0, \cdots, m\,; \; N \in {\bf Z}\, ).
\label{poleone}
\ee
In the total amplitude (\ref{integralrep}) some of these poles may
not appear due to cancellations in the sum over $l, \; m, \; n$.
However, except for special cases, the poles (\ref{poleone}) in
general survive after the summation in eq.\ (\ref{integralrep}).
The amplitude (\ref{integralrep}) with eq.\ (\ref{poleone}) has a
quite different form from that of the sphere
topology \cite{DFKU}, \cite{SATACO}.
\par
In the case (ii) we find that $I_{lmn}$ vanishes when $m$ is
non-zero. Therefore we only have to consider the case $m=0$,
which is given by
\ba
I_{l 0 n}
&\!\!\! = &\!\!\! (-1)^l \left[ {\pi \over \Gamma(1+\rho)}
\right]^s {1 \over \Gamma(1-s\rho)} \nonu
&\!\!\! &\!\!\! \times \prod_{k=1}^l {1 \over \sin (\pi k \rho)} \,
  \prod_{k=1}^n {1 \over \sin (\pi k \rho)} \,
  \prod_{k=1}^s {1 \over s-k+(1+a)\rho^{-1}}.
\label{casetwo}
\ea
It has a finite number of poles at
\be
p_1 = {1 \over 2}(k-1)\alpha_+ + {1 \over 2}\alpha_-
\quad \ \ \ (\, k = 1, 2, \cdots, s \, ).
\label{poletwo}
\ee
The pole structure (\ref{poletwo}) is independent of $l, \; n$.
As a consequence the sum in eq.\ (\ref{integralrep}) identically
vanishes when $s$ is an odd integer. However, for an even integer
$s$ the total amplitude (\ref{integralrep}) does not vanish in
general. This is in contrast to the case of sphere topology,
in which amplitudes in the kinematical region (ii)
identically vanish \cite{DFKU}.
\par
\figuretwo
Vanishing of the amplitude $I_{lmn}$ for $m \not= 0$ is
a special case of more general vanishing amplitudes.
In ref.\ \cite{BEKU} it was argued that only non-vanishing
$N$-point tachyon amplitudes are of the type
$(+, +, \cdots, +, -, -, \cdots, -)$ shown in Fig.\ 2 (a),
where the signs denote the chiralities of the tachyons.
The amplitude vanishes if it has two or more groups of tachyons of
negative chirality, each of which are between tachyons of positive
chirality. An example of such amplitudes is shown in Fig.\ 2 (b).
\par
%
%
\newsection{Factorizations}
Here we examine the origin of pole singularities of the amplitudes
obtained in the previous section. In the case of the $c=1$ theory
on the sphere, the momentum singularities can be understood by
factorizations with the tachyon and the discrete states as
intermediate
states \cite{POLYAKOVSELF}, \cite{SATAFACT}, \cite{DFKU}.
We will see that the singularities of the disk amplitudes can also
be understood by factorizations. The factorization analysis can
be applied to the $c<1$ case as well as to the $c=1$ case.
\par
After the zero-mode integrations, the
amplitude (\ref{zeromodeint}) can be regarded as a
$(3 + s)$-point function of three tachyon operators
$O_{\rm o}(p_1),\; O_{\rm o}(p_2),\; O_{\rm o}(p_3)$ and
$s$ cosmological operators $O_{\rm o}(p \! = \! 0)$ with the
zero modes omitted. These operators are
integrated along the boundary of the disk. The term $I_{lmn}$ in
eq.\ (\ref{integralrep}) is a contribution from the integration
region shown in Fig.\ 1. When some of these $3+s$ operators
approach one another, short distance singularities arise by the
OPE, which give rise to singularities in momenta after integrations
over positions of the operators. In contrast to the case of the
sphere, only neighboring operators on the boundary can approach
one another. The momentum singularities appear at the momentum for
which the operator produced by the OPE is a physical operator,
i.e. the tachyon or the discrete states.
\par
First we examine the case (i). Consider the integration region in
which the tachyon 1, $k$ out of $m$ cosmological operators, and $q$
out of $n$ cosmological operators approach one another.
It gives a factorization of the amplitude shown in Fig.\ 3.
The momentum singularities can arise when the two-momentum of the
intermediate state $p^\mu = p_1^\mu + (k+q) \alpha^\mu$
($\alpha^\mu = (0, -i \alpha_+)$) is that of the tachyon or the
discrete states. For the momentum $p^\mu$ to be that of the
tachyon, the momentum of the tachyon $1$ must be
\be
p_1 = - {1 \over 2}(q+k+1)\alpha_+ + {1 \over 2} \alpha_-
\label{tachpole}
\ee
and the intermediate tachyon has the negative chirality.
On the other hand, for $p^\mu$ to be the momentum of the discrete
state, the momentum of the tachyon $1$ must be
\be
p_1 = - {1 \over 2}(q+k+1)\alpha_+ + {1 \over 2}(t+1) \alpha_-,
\label{discpole}
\ee
where $t$ is an arbitrary positive integer. The intermediate
discrete state is $(r=q+k, t)$ of A-type in eq.\ (\ref{discrete}).
For $q=n$ and $k=0, 1, 2, \cdots, m$ these values of momenta
(\ref{tachpole}), (\ref{discpole}) coincide with the positions of
the singularities (\ref{poleone}) for $N \! \geq \!1$.
The rest of the singularities $N < 1$ in eq.\ (\ref{poleone}) is
explained in a similar way by the factorization with the
tachyons $1$ and $2$ are interchanged in the above case.
\figurethree
\par
One expects that the amplitude has singularities also at
the momenta (\ref{tachpole}) and (\ref{discpole}) for $q < n$.
However, they do not appear in the amplitude (\ref{caseone}).
This can be understood as follows. First consider the case in
which the intermediate state is the tachyon. The intermediate
tachyon is of negative chirality as noted above.
When $q <n$, this intermediate
tachyon lies between operators of positive chiralities in the
right side blob of Fig.\ 3. From the discussion at the end of
the previous section such an amplitude vanishes and therefore
there arise no singularity for $q < n$. The absence of the
singularities at the momentum (\ref{discpole}) for $q < n$
suggests that amplitudes also vanish if there exists a discrete
state of A-type between operators of positive chiralities
in addition to negative chirality tachyons between operators of
positive chiralities. Accepting the vanishing of
these amplitudes, the singularities (\ref{poleone}) are
completely explained by the factorizations.
\par
The singularities (\ref{poletwo}) in the case (ii) can be
understood by factorizations in the same way. The integration
region in which the tachyon 1, $q$ out of $n$ cosmological
operators approach one another gives a factorization of
the amplitude shown in Fig.\ 4. From the kinematics, only the
tachyon of positive chirality is allowed as an intermediate state.
For the intermediate momentum $p^\mu$ to be that of the tachyon,
the momentum of the tachyon $1$ must be
\be
p_1 = {1 \over 2}(q-1)\alpha_+ + {1 \over 2} \alpha_-.
\label{tachpoletwo}
\ee
These values of momenta coincide with the positions of
the singularities (\ref{poletwo}) for $k = 1, 2, \cdots, n$.
The rest of the singularities $k = n+1, n+2, \cdots, n+l \, (=s)$
in eq.\ (\ref{poletwo}) is explained by the factorization with the
tachyons $1$ and $2$ are interchanged in the above case.
\figurefour
\par
Thus all structures of pole singularities, which we have obtained
in the previous section, have been understood by the factorizations
with the tachyon and the discrete states as intermediate states.
\par
\end{document}